# Increasing Modal Instabilities Threshold Through Bidirectional Pumping In High Power Fiber Amplifiers

Zeinab Sanjabi Eznaveh, Gisela Lopez-Galmiche, and Rodrigo Amezcua Correa

*Abstract*—In this letter, the impact of pump configuration on the modal instabilities (MI) threshold of high power fiber amplifiers is investigated. Using a time dependent computer model based on the optical beam propagation method (BPM), the effect of bidirectional pumping on the MI threshold of high power fiber amplifiers is explored and compared to that of a copumped configuration. Simulation results indicate that the bidirectional pumping increases the instabilities threshold by a factor of 30% with respect to the copumped scheme. We attribute this threshold enhancement to a weaker strength of the temperature induced refractive index grating due to a more homogeneous pump profile along the fiber length. This technique demonstrates that the engineering pump scheme is a promising approach leading to an appreciable threshold enhancement of fiber amplifiers designed for high power applications.

*Index Terms*— modal instabilities (MI), fiber amplifiers, beam propagation method (BPM), bidirectional pumping, refractive index grating

## I. INTRODUCTION

Fiber lasers have demonstrated an impressive development in recent years due to their remarkable power scalability. However, due to the high optical intensity in the core, the performance of the high power fiber lasers is subject to several detriments nonlinear processes such as four-wave mixing (FWM), self-phase modulation (SPM), stimulated Brillouin scattering (SBS) and stimulated Raman scattering (SRS) [1]. To mitigate nonlinear effects large mode area (LMA) fibers that exhibit a mode field diameter (MFD) larger than 50um have been developed. However, in the other hand, large core sizes allow the propagation of higher order modes (HOMs), which might potentially lead to degradation of the output beam quality. Recently, thermal modal instability (MI) which is defined as the kHz fluctuations of the output beam of a fiber amplifier has been introduced as one of the most severe limitations on the power scalability of LMA ytterbium-doped (Yb-doped) fiber amplifiers [2]-[4]. Several theoretical and experimental investigations reported to date have proven that the energy transfer between modes is caused by a thermally-induced index grating due to quantum defect heating of the active fiber [5]–[14]. These observations also revealed the MI threshold and its oscillation frequency are influenced by fiber and amplifier parameters such as core and cladding size, core numerical aperture (NA), wavelength, seed power, quality of the mode excitation, the temporal and the spectral properties of signal and pump, etc. [3].

In this letter we examine how the instabilities threshold is influenced by pump distribution profile. To do so, we compare the impact of bidirectional pumping on the MI threshold in an Yb-doped fiber amplifier with an amplifier under copumped condition. To accomplish this, we apply BPM which considers: laser gain, thermal lensing and spatial evolution of the signal field in the absence of frequency shifted modes similar to the models presented in [11]-[12], [15].

Through numerical simulations, it is shown that bidirectional pump configuration can effectively be used to increase the MI threshold by a factor of 30% as compared to the copumped configuration. We attribute this threshold enhancement to a weaker strength of the induced refractive index grating in the fiber amplifier pumped under bidirectional condition. Ultimately, we conclude that engineering pump scheme is a promising technique that can be used to considerably enhance the threshold of MI in high power fiber lasers and amplifiers.

## II. THEORETICAL MODEL

To model the dynamic behavior of MI in high power fiber amplifiers we apply a BPM taking into account laser gain and refractive index perturbations, a heating model based on pump and signal intensity profiles and quantum defect heat source and a time-dependent temperature solver. Here, we summarizes the computational algorithm of the BPM used to model the dynamic behavior of the MI in high power fiber amplifiers. The mathematical equations and model description of BPM have been previously described in multiple reports [12], [15]-[16].

This work was supported by U.S. Army Research Office and High Energy Laser Joint Technology Office and under Multidisciplinary Research Initiative W911NF-12-1-0450.

Z. Sanjabi Eznaveh is with the Collage of Optics and Photonics, CREOL, University of Central Florida, Orlando, FL 32816 USA, (zahoora@knights.ucf.edu).

G. Lopez-Galmiche is with CREOL, University of Central Florida, Orlando, FL 32816 USA, (gisela.lopezgalmiche@ucf.edu).

R. Amezcua-Correa is with CREOL, University of Central Florida, Orlando, FL 32816 USA, (r.amezcua@creol.ucf.edu).

Firstly, based on the fiber parameters, the normalized initial electric fields lunched into the fiber are calculated. (Fig.1-b, I). The excited state population which creates an inversion profile, is then computed in the doped section of the fiber using the laser rate equations (Fig.1-b, II). The Inversion profile leads to the heat profile due to the absorbed pump power and quantum defect - defined as the energy difference between the pump and signal wavelengths. Heat profile creates a temperature profile across the fiber surface (Fig.1-b, III) which in turn induces a refractive index grating in the fiber caused by the thermo-optic effect (Fig.1-b, IV). Due to the successive variations of the index profile along the fiber length, one should update the refractive index of the fiber and the related physical parameters such as laser gain and the phase of the electric fields at each length step, $\Delta L$. As a result, the signal and pump fields experience a novel waveguide medium as propagating towards the end facet of the fiber. So far, all the processes explained occur at just one time discretization of a dynamic behavior. To complete the time dependent model of the MI a repetitive procedure should be accomplished over a ms time duration to fully capture the dynamic performance of the MI.

**Table 1. Amplifier Parameters**

| Parameters | Value |
|---|---|
| Core diameter | 50 μm |
| Pump cladding diameter | 250 μm |
| Outer cladding diameter | 500 μm |
| Core refractive index | 1.45 |
| Numerical aperture | 0.03 |
| Fiber length | 1.6 m |
| Signal wavelength | 1064 nm |
| Pump wavelength | 977 nm |
| Signal $LP_{01}$ power | 28.5 W |
| Signal $LP_{11}$ power | 1.5 W |
| $Yb^{+3}$ doping concentration | $6 \times 10^{25}$ 1/m$^3$ |
| Signal emission cross-section | $3.58 \times 10^{-25}$ m$^2$ [12] |
| Signal absorption cross-section | $6 \times 10^{-27}$ m$^2$ [12] |
| Pump emission cross-section | $1.87 \times 10^{-24}$ m$^2$ [12] |
| Pump absorption cross-section | $1.53 \times 10^{-24}$ m$^2$ [12] |
| Upper state lifetime | 850μs |
| Thermo-optic coefficient | $1.29 \times 10^{-5}$ K$^{-1}$ |
| Thermal conductivity | 1.38 W/(m-K) |
| Heat capacity | 703 J/kg-K |
| Mass density | 2200 m$^3$ |

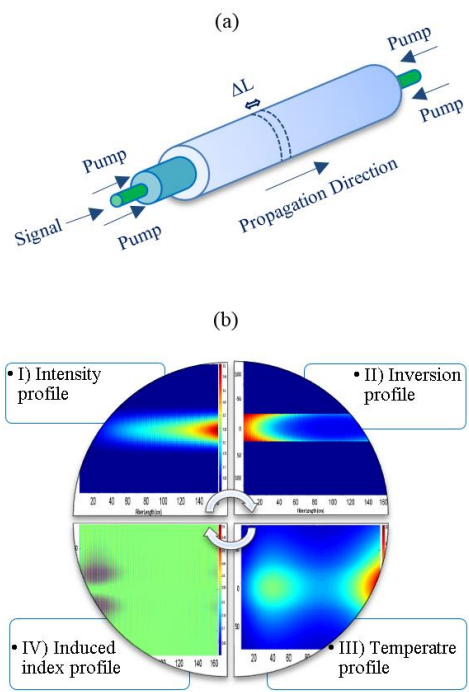

**Fig.1**. Schematic representation of the double clad Yb-doped fiber amplifier pumped bi-directionally (a), dynamic model of the MI in high power fiber amplifiers under bidirectional pump condition (b)

### III. SIMULATION RESULTS AND DISCUSSION

For the simulations, we consider a 1.6m long, LMA cladding-pumped step-index Yb-doped fiber, cooled under Dirichlet boundary conditions. The amplifier parameters which have been considered in the previous MI studies [11]-[12] are listed in table 1:

#### A. Copumped Amplifier

As the first set of simulations, we consider launched co-pumped powers of 760W, 920W and 1100W corresponding to below, around and above MI threshold. The total seed power is 30W from which 95% is in the fundamental mode and 5% in the x-polarized first HOM ($LP_{11}$) with no frequency shift between the modes.

Figure 2(a) illustrates the evolution of the optical powers along the propagation direction for the pump power of 760W and the accompanying movie (Media1) shows the temporal evolution of the intensity distributions at the output end of the fiber amplifier within the core over 80ms time duration. It is evident from the results that during the first 10-20ms of the amplifier operation, the power oscillates back and forth between the fundamental mode (the green plot) and the first higher order mode (the red plot) related to the abrupt turn on of the pump power. Once the amplifier approaches the quasi-equilibrium region, since the pump is not sufficiently high to onset the energy transfer between the interfering modes, the signal power is largely in the fundamental mode. This power regime is referred to the region one or below threshold.

By increasing the pump power to 920W the amplifier enters a transient region where the two interfering modes transfer energy more significantly. Figure 2(b) and Media2 are the corresponding results of this region.

Further increase of the input pump power leads to a very rapid transfer of energy between the interfering modes. In this region the amplifier shows a "chaotic" behavior and there is no sign of stability or periodic oscillations. Figure 2(c) and Media3 display the corresponding plots for a pump power of 1100W.

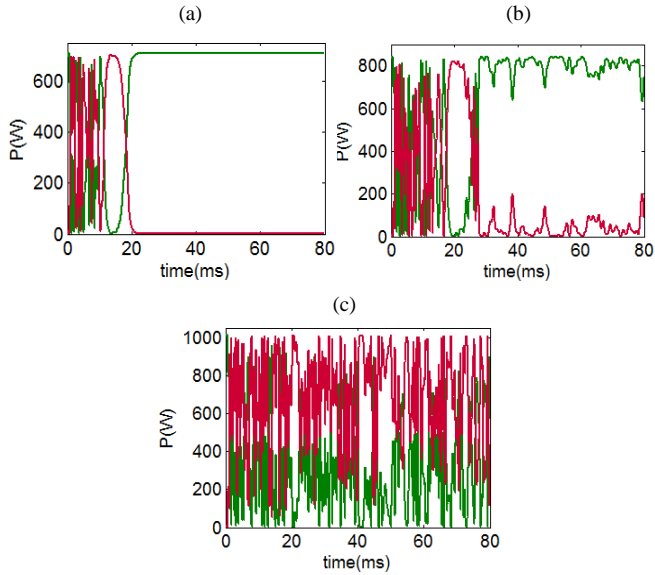

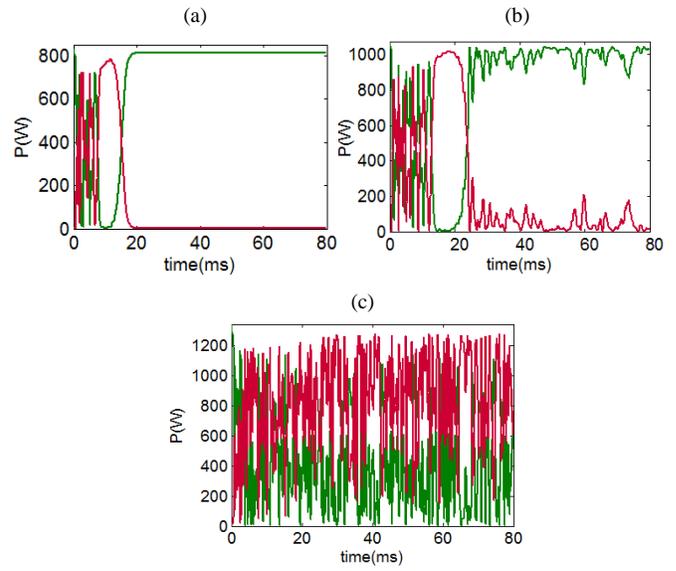

Fig.2. Power evolution of the fundamental mode (Green) and the first HOM (Red) over 80ms at the output end of the fiber amplifier at the copumped power of (a) 760W, (b) 920W, and (c) 1100 W. The corresponding movies (Media1, Media2, and Media3) show the temporal variations of the modal powers along the propagation direction.

Fig.3. Power evolution of the fundamental mode (Green) and the first higher order mode (Red) over 80ms at the output end of the fiber amplifier under bi-directional pumping for pump power of (a) 905W, (b) 1200W, and (c) 1500 W. The corresponding movies (Media4, Media5, and Media6) show the temporal variation in modal powers along the propagation direction.

By convention, we defined the MI threshold to be the power level at which the ratio of the HOM content to the total signal power is 5%. For the amplifier under test, the MI threshold was estimated to be 840W of the output power corresponding to a launched pump power of 905W.

*B. Bidirectional Pumping*

To investigate the behavior of the fiber amplifier under bidirectional pump configuration, we assume an amplifier with the same parameters defined in the previous section but the input pump power equally divided between co and counter pumps. The total launched pump powers are chosen to be 905W, 1200W and 1500W with a signal seed power of 30W.

Figure 3 illustrates the evolution of the optical powers along the propagation direction and the accompanying movies (Media4, Media5 and Media6) show the temporal evolution of the intensity distributions within the core at the output end of the fiber amplifier over 80ms time duration. From Fig.3 (a) it can be deduced that the total power of 905W which was estimated as the MI threshold in the case of copumped amplifier, depicts a stable behavior for the case of bidirectional pumping.

Simulation results show that the MI threshold in the case of bidirectional pumping is at about 1025W of the output signal power -corresponding to 1170W of total input pump power which implies an improvement of the MI threshold by a factor of around 30% as compared to the copumped case. In addition, further investigations confirms this power ratio (half forward-half backward) leads to the maximum achievable MI threshold in the case of bidirectional pumped amplifiers. Therefore, by engineering the pump scheme a significant improvement in the MI threshold in high power fiber amplifiers can be achieved.

At this point it is worth mentioning that in the dynamic model of a fiber amplifier under bidirectional pump condition, the initial power in the counter-pumped direction is not completely stable (in our case the maximum change was in the order of less than 10% of the designated counter-pumped input power) which in fact, underestimates the threshold of MI under bidirectional pump condition due to pump power fluctuations. To be more exact, it means, the actual improvement of the MI threshold of the amplifier under test, pumped bidirectional ly is even more than 30% with respect to the copumped condition which further manifest the advantage of the current engineered pump configuration.

Studies show the saturation effects are in fact responsible for higher obtained instabilities threshold in the case of bidirectional pumping. To more detailed explanations, as light is propagating along the fiber, spatial-temporal temperature variations arising from quantum defect heating through the thermo-optic effect induces a long period refractive index grating. Due to the long thermal diffusion time in silica (0.7ms approximately for the current core radius), this thermally induced long period grating (LPG) is out of phase with respect to the signal intensity pattern which therefore results in an efficient energy transfer between interfering modes. However, in an efficient fiber amplifier transverse spatial hole-burning that depletes the population inversion should be mainly taken into account [17]. Based on this fact, in a bidirectional pumped amplifier where the net pump power maintains an efficiently moderate value almost along the entire length of the fiber, this effect is strong enough to influence the depletion of inversion population along the propagation direction and thus it changes the heat deposition profile along the fiber and gives rise to a weaker strength of the induced index grating to transfer energy between the modes and consequently a higher MI threshold is achievable. This is illustrated more clearly in Fig.(4) which represents the temperature induced index grating at the central line of the core, for both cases of copumped (F) and bidirectional pumped (Bi) configurations at their thresholds after 40ms and 80ms.

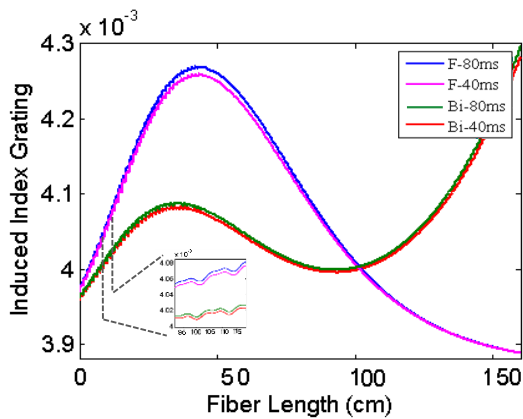

**Fig.4**. Temperature induced index gratings at the central line of the core of the fiber for two cases of copumped (F) and bidirectional pumped (Bi) amplifiers at their thresholds after 40ms (bottom plot) and 80ms (top plot) time duration.

Looking at Fig.4 (Close-up) reveals that under bidirectional pump condition the maximum strength of the index change due to temperature (between points of 10.1cm and 11.1cm of the fiber length) is $17e^{-3}$ whilst this value under copumped condition (between points of 10cm and 10.9cm of the fiber length) is $11e^{-3}$ which results in a 35% degradation of the strength of the induced grating which consequently, as explained earlier, leads to an increase of around 30% of the TMI threshold.

## IV. CONCLUSION

We investigated the influence of pump profile on the modal instabilities threshold in high power fiber amplifiers. To fulfill this, the effect of bidirectional pumping on the mode instabilities threshold was investigated and compared to the copumped condition through a dynamic model of beam propagation method. The simulation results displayed ~30% increase in the modal instabilities threshold under bidirectional pumping relative to the copumped amplifier. Studies show the temperature induced index grating caused by heat deposition while taking saturation effects into account, is responsible for this threshold enhancement. Ultimately, we conclude that pump engineering is a promising technique to considerably increase the threshold of modal instabilities in high power fiber lasers and amplifiers.

## ACKNOWLEDGMENTS

This work was supported by U.S. Army Research Office and High Energy Laser Joint Technology Office and under Multidisciplinary Research Initiative W911NF-12-1-0450. We thank STOKES Advanced Research Computing Center (ARCC) at the University of Central Florida for providing computational resources for simulation works. G.G. acknowledges support from CONACYT.

## REFERENCES

[1] G. P. Agrawal, *Nonlinear Fiber Optics*, 4th Edition. NY, USA: Academic Press, 2006.
[2] F. Stutzki, et al., "High average power large-pitch fiber amplifier with robust single-mode operation," Opt. Lett. vol. 36, no. 5, pp. 689–691, Mar. 2011.
[3] T. Eidam, et al., "Experimental observations of the threshold-like onset of mode instabilities in high power fiber amplifiers," Opt. Exp. vol. 19, no. 14, pp. 13218–13224, Jun. 2011.
[4] K. R. Hansen, T. T. Alkeskjold, J. Broeng, and J. Lægsgaard, "Theoretical analysis of mode instability in high-power fiber amplifiers," Opt. Exp. Vol. 21, no. 2, pp. 1944-1971, Jan. 2013.
[5] F. Stutzki, "High-speed modal decomposition of mode instabilities in high-power fiber lasers," Opt. Lett. vol. 36, no. 23, pp. 4572–4574, Dec. 2011.
[6] A. V. Smith and J. J. Smith, "Mode instability in high power fiber amplifiers," Opt. Exp. vol. **19**, no. 11, pp. 10180–10192, May. 2011.
[7] A. V. Smith, and J. J. Smith, "Influence of pump and seed modulation on the mode instability thresholds of fiber amplifiers," Opt. Exp. vol. 20, no. 22, pp. 24545–24558, Oct. 2012.
[8] C. Jauregui, et al., "Temperature-induced index gratings and their impact on mode instabilities in high-power fiber laser systems," Opt. Exp. vol. 20, no. 1, pp. 440–451, Dec. 2012.
[9] K. R. Hansen, T. T. Alkeskjold, J. Broeng, and J. Lægsgaard, "Thermally induced mode coupling in rare-earth doped fiber amplifiers," Opt. Lett. vol. 37, no. 12, pp. 2382–2384, Jun. 2012.
[10] L. Dong, "Stimulated thermal Rayleigh scattering in optical fibers," Opt. Exp. vol. 21, no. 3, pp. 2642–2656, Jan. 2013.
[11] S. Naderi, I. Dajani, T. Madden, and C. Robin, "Investigations of modal instabilities in fiber amplifiers through detailed numerical simulations,' Opt. Exp. vol. 21, no. 13, pp. 16111-16129, Jun. 2013.
[12] B. Ward, "Modeling of transient modal instability in fiber amplifiers," Opt. Exp. vol. 21, no. 10, pp. 12053–12067, May. 2013.
[13] A. V. Smith, and J. J. Smith, "Increasing mode instability thresholds of fiber amplifiers by gain saturation," Opt. Exp. vol. 21, no. 13, pp. 15168–15182, Jun. 2013.
[14] C. Jauregui, et al., "Passive mitigation strategies for mode instabilities in high-power fiber laser systems," Opt. Exp. vol.21, no.16, pp. 19375-19386, Aug. 2013.
[15] A. V. Smith and J. J. Smith, "Steady-periodic method for modeling mode instability in fiber amplifiers," Opt. Exp. vol.21, no.3, pp. 19375-19386, Jan. 2013.
[16] M. D. Feit and J. A. Fleck, "Computation of mode properties in optical fiber waveguides by a propagating beam method," Appl. Opt. vol. 19, no. 7, pp. 1154-1164, Apr. 1980.
[17] A. V. Smith and J. J. Smith, "Overview of a steady-periodic model of modal instability in fiber amplifiers," J. of selected topics in quantum electronics vol. 20, no.6, Dec. 2014.